\documentclass[sn-basic]{sn-jnl}

\usepackage{makecell}	
\usepackage{natbib}	
\usepackage{tabularx}	

\theoremstyle{thmstyleone}%
\theoremstyle{thmstyletwo}%

\theoremstyle{thmstylethree}%

\begin{document}

\title{MOPRD: A multidisciplinary open peer review dataset}


\author[1,3]{\fnm{Jialiang} \sur{Lin}}\email{me@linjialiang.net}

\author[1,3]{\fnm{Jiaxin} \sur{Song}}\email{songjiaxin@stu.xmu.edu.cn}

\author[2]{\fnm{Zhangping} \sur{Zhou}}\email{zhouzp@xmu.edu.cn}

\author[1,3]{\fnm{Yidong} \sur{Chen}}\email{ydchen@xmu.edu.cn}

\author*[1,3]{\fnm{Xiaodong} \sur{Shi}}\email{mandel@xmu.edu.cn}

\affil[1]{\orgdiv{School of Informatics}, \orgname{Xiamen University}, \orgaddress{\city{Xiamen}, \country{China}}}

\affil[2]{\orgdiv{College of Foreign Languages and Cultures}, \orgname{Xiamen University}, \orgaddress{\city{Xiamen}, \country{China}}}

\affil[3]{\orgdiv{Key Laboratory of Digital Protection and Intelligent Processing of Intangible Cultural Heritage of Fujian and Taiwan}, \orgname{Ministry of Culture and Tourism}, \orgaddress{\city{Xiamen}, \country{China}}}


\abstract{Open peer review is a growing trend in academic publications. Public access to peer review data can benefit both the academic and publishing communities. It also serves as a great support to studies on review comment generation and further to the realization of automated scholarly paper review. However, most of the existing peer review datasets do not provide data that cover the whole peer review process. Apart from this, their data are not diversified enough as the data are mainly collected from the field of computer science. These two drawbacks of the currently available peer review datasets need to be addressed to unlock more opportunities for related studies. In response, we construct MOPRD, a multidisciplinary open peer review dataset. This dataset consists of paper metadata, multiple version manuscripts, review comments, meta-reviews, author's rebuttal letters, and editorial decisions. Moreover, we propose a modular guided review comment generation method based on MOPRD. Experiments show that our method delivers better performance as indicated by both automatic metrics and human evaluation. We also explore other potential applications of MOPRD, including meta-review generation, editorial decision prediction, author rebuttal generation, and scientometric analysis. MOPRD is a strong endorsement for further studies in peer review-related research and other applications.}

\keywords{Open peer review, Review comment generation, Dataset applications, Digital libraries}



\maketitle

\section{Introduction}
\label{sec:introduction}

Traditional peer review~\citep{ware-stm-2015} is based on an anonymous mechanism~\citep{nobarany-understanding-2017}. With this anonymous mechanism, the reviewers' identities and the review reports are not disclosed to the public. However, the anonymous mechanism might in turn compromise the peer review process~\citep{morrison-case-2006,khan-is-2010,laine-annals-2017}. As a better solution to the flawed anonymous mechanism, open peer review, a more transparent mechanism, has been introduced to academic publications~\citep{ford-defining-2013}. Open peer review is proved to attract review comments of higher quality in terms of length, helpfulness, and good manners~\citep{bornmann-closed-2012,walsh-open-2000}. In the 1990s, open peer review was first adopted by the \textit{BMJ}~\citep{vanrooyen-effect-1998,vanrooyen-effect-1999}. With more and more journals following the example, this mechanism is gradually recognized by the academic and publishing communities. As of December 2019, a total of 617 journals have taken up the practice of open peer review~\citep{wolfram-open-2020}. These include some leading journals like \textit{Nature}, which offers open peer review as an option.

Open peer review unfreezes access to a large amount of peer review data, unlocking great opportunities for computer-aid peer review and further in the realization of automated scholarly paper review (ASPR)~\citep{lin-automated-2023}. These data, especially structured data, are of great importance to artificial intelligence (AI) research~\citep{lin-automatic-2022}. However, open peer review data are scattered across different platforms in various formats. To tap into the full value of open peer review, these data need to be collected and processed into datasets of structured data for further use. As more and more journals and conferences release their review reports, efforts have been made to construct peer review datasets. Examples of popular open peer review datasets include PeerRead~\citep{kang-dataset-2018} and ASAP-Review~\citep{yuan-can-2022}. These peer review datasets have been used for different research purposes and practical applications, e.g.\ acceptance prediction, score prediction, and review comment generation. However, there exist some limitations and inadequacies as follows in the construction of open peer review datasets.


Firstly, most of the currently existing datasets do not contain data that cover the whole peer review process. A complete peer review dataset should include the initial submissions, review comments, meta-reviews, revisions, author's rebuttal letters, and editorial decisions. Some existing datasets are built on the NeurIPS (previously named NIPS) Proceedings.\footnote{https://proceedings.neurips.cc/.} NeurIPS provides peer review data in text format to the public, but the initial submissions, based on which the review comments are written, are not available. Without the initial submissions, before-and-after studies can not be conducted between the initial submission and the published version of a paper for further understanding of peer review comments.

Secondly, a large proportion of presently available datasets do not include papers across different disciplines. The majority of the publicly available open peer review datasets are datasets of computer science. This is because open peer review data in computer science are most easily accessible compared to those in other disciplines. OpenReview\footnote{https://openreview.net/about.} is one example of data sources for open peer review in computer science. Apart from this, the construction of datasets is also a task in the field of computer science. Most of the researchers building the datasets are in computer science, and this also explains why computer science papers are more often chosen for the creation of open peer review datasets. Single-discipline datasets fall short in data diversity, which renders them incapable of supporting multidisciplinary research. It is also difficult to reuse models trained with single-discipline data for tasks involved in other disciplines.

In response to the above-mentioned problems that trouble current open peer review datasets, we design special crawlers to obtain complete peer review history data with native and auto-labeling labels from journals across different disciplines. Our work and main contributions are:

\begin{itemize}

\item We construct and publish the Multidisciplinary Open Peer Review Dataset (MOPRD). MOPRD is composed of paper metadata, manuscripts of the initial submission and following revisions, review comments, meta-reviews, author's rebuttal letters, and editorial decisions of papers across various disciplines. These data empower MOPRD to be an open peer review dataset of great completeness and diversity;

\item We design a modular guided review comment generation method based on MOPRD. This method is capable of processing long-text input and generating well-structured review comment output;

\item We explore other applications of MOPRD for tasks including meta-review generation, editorial decision prediction, author's rebuttal generation, and scientometric analysis.

\end{itemize}

\section{Related work}
\label{sec:relatedwork}

Currently, there are many peer review datasets available for various purposes. These datasets can be broadly classified into three types by their main applications, which are Prediction, Analysis, and Generation. Table~\ref{tab:peer-review-dataset} illustrates these three types of peer review datasets with their domains and applications.

\begin{table}[htb]
    \centering
	\caption{Related peer review datasets}
	\label{tab:peer-review-dataset}
	\resizebox{0.96\textwidth}{!}{
		\begin{tabularx}{\textwidth}{|m{0.65cm}<{\centering}|m{4.3cm}|m{1.2cm}|m{4cm}|}
			\hline
			Type  &  Name  &  Domain  &  Application    \\
			\hline
			P  &  PeerRead~\citep{kang-dataset-2018}          &  AI  &  Acceptance prediction, score prediction  \\
			\hline
			P  &  \makecell[l]{ACL-18 Numerical \\ \citep{gao-does-2019}}      &  NLP  &  After-rebuttal score changing prediction  \\
			\hline
			P  &  \makecell[l]{Interspeech 2019 Submission \\ \citep{stappen-uncertainty-2020}}  &  Speech  &  Acceptance prediction, score prediction  \\
			\hline
			A  &  AMPERE~\citep{hua-argument-2019}            &  AI  &  Review discourse analysis  \\
			\hline
			A  &  \makecell[l]{CiteTracked \\ \citep{plank-citetracked-2019}}  &  AI  &  Citation prediction  \\
			\hline
			A  &  COMPARE~\citep{singh-compare-2021}          &  AI  &  Discussion sentence comparison  \\
			\hline
			A  &  Dataset of \citet{matsui-impact-2021}        &  \makecell[l]{BIO, CS, \\ ENV}  &  Peer review process analysis  \\
			\hline
			A  &  ReAct~\citep{choudhary-react-2021}          &  AI  &  Review requirement analysis  \\
			\hline
			A  &  Dataset of \citet{ghosal-peer-2022}          &  AI  &  Review comment analysis  \\
			\hline
			G  &  ASAP-Review~\citep{yuan-can-2022}           &  AI  &  Review comment generation  \\
			\hline
			G  &  MReD~\citep{shen-mred-2022}                 &  AI  &  Meta-review generation  \\
			\hline
		\end{tabularx}

	}
	\begin{tablenotes}
		\item \footnotesize{In column Type, P is short for Prediction, A for Analysis, and G for Generation}
	\end{tablenotes}
\end{table}

The first type is datasets mainly used for acceptance prediction and score prediction. PeerRead~\citep{kang-dataset-2018} is the first large-scale peer review dataset. It consists of full text papers from ACL, ICLR, and NIPS and the editorial accept/reject decisions on them. Most of the papers in PeerRead are collected together with their review comments, and some of them also have aspect specific scores such as appropriateness, clarity, and originality. This dataset unlocks great opportunities for research on peer review and has already enabled several peer review studies. ACL-18 Numerical Peer Review Dataset~\citep{gao-does-2019} is a collection of review comments, author's rebuttal as well as scores given by reviewers before and after the rebuttal phase. This dataset opens opportunities for studies on the efficacy of rebuttal. Interspeech 2019 Submission~\citep{stappen-uncertainty-2020} is constructed by the organizers of the conference. It contains the original text of all manuscripts submitted to Interspeech and their corresponding review comments, scores, and decisions. What makes this dataset particularly valuable is its collection of a large number of rejected manuscripts, but unfortunately the dataset is not publicly available yet.

The second type is datasets primarily employed to analyze different aspects of peer review process. \citet{hua-argument-2019} create AMPERE with 10,386 labeled propositions from 400 review comments. This dataset is built with a focus on reviewers' argumentative propositions and their types. The authors train proposition segmentation and classification models with this dataset to analyze the patterns and evaluate the utilities of argumentative propositions. Built by \citet{plank-citetracked-2019}, CiteTracked has its focus on the impact of research papers. It is composed of papers of NeurIPS, their review comments, and the number of citations. CiteTracked is constructed for the task of citation count prediction on the basis of review comments. \citet{singh-compare-2021} construct COMPARE, a dataset of comparative discussion sentences collected from review comments, in their research to achieve more efficient peer review. \citet{matsui-impact-2021} analyze the peer review process based on data collected from PeerJ. Papers of biology, computer science, and environment are studied with traditional machine learning technologies to understand the impact of peer review process on papers, including the acceptance timeline, potential contributions, manuscript modifications. ReAct~\citep{choudhary-react-2021} is a review comment dataset with classified review comments used to understand the requirements of the reviewers to the authors. It is constructed along with the proposal of a new task of determining whether the review comments contain requirements for the authors to act on. \citet{ghosal-peer-2022} build the first multi-layered peer review dataset with 17k sentences from 1,199 review comments. The review comments are labeled across four layers to study the corresponding sections in the reviewed paper, the aspects of the review text, the role of the review text, and the importance of the review statement. Based on these four layers, four new tasks are introduced in the understanding of review comments, which can be used as indicators for the evaluation of peer review performance.

The third type is datasets principally designed for comment generation. With 28,119 review comments, ASAP-Review~\citep{yuan-can-2022} is the most extensive dataset of its kind in the field of computer science. It also provides sentence-level aspect labels, such as Summary, Originality, and Clarity, and has been used for a series of studies for review comment generation. \citet{shen-mred-2022} construct MReD with a total of 45,929 sentences from 7,089 labeled meta-reviews. Based on this dataset, structurally controllable extractive and abstractive text summarization models are proposed.

Compared to the datasets mentioned above, to our best knowledge, MOPRD is by far the largest multidisciplinary peer review dataset with complete peer review history. With this feature, MOPRD is capable of different applications. Thus, studies based on MOPRD can be more universal and more generalized.

\section{MOPRD}

In this section, we elaborate on the data source selection, construction process, and properties of the Multidisciplinary Open Peer Review Dataset (MOPRD).

\subsection{Data source}

According to the study of \citet{wolfram-open-2020} on open peer review, as of 2019, 617 journals from 38 publishers have been practicing open peer review. Ideally, the best practice for constructing an open peer review dataset would be collecting all the data of open peer review from all these journals in an exhaustive manner. More importantly, in order to keep up with the latest publications, data collection should be a continuous process. Practically, however, such a high-maintenance dataset would be too demanding to create. One major impediment is that most of the data are unstructured and scattered on different web pages across different journals. Collecting all these data requires complex and costly organization, which is difficult for this project alone to achieve. An exhaustive collection of all open peer review data will be our long-term goal with crowd-sourcing being one of the possible solutions. In the face of such an arduous task, we take the initial step to start small. In this project, careful filtering is made in the selection of our data source among journals that are currently practicing open peer review.

Taking into consideration scopes, influence, data integrity, data format, and license, PeerJ is selected as the data source for the construction of MOPRD. PeerJ is a large general academic publisher of up to seven journals: \textit{PeerJ Analytical Chemistry}, \textit{PeerJ Computer Science}, \textit{PeerJ Inorganic Chemistry}, \textit{PeerJ - Life and Environment}, \textit{PeerJ Materials Science}, \textit{PeerJ Organic Chemistry}, and \textit{PeerJ Physical Chemistry}. These seven journals cover a wide range of research disciplines, e.g.\ biology, chemistry, computer science, environment, and medicine.

One feature that distinguishes PeerJ journals from other journals is its encouragement of review transparency. Authors that submit papers to PeerJ are given the option to present the complete peer review history along with the final published paper to readers. For each paper, this peer review history includes the initial submission, revisions, review comments, meta-reviews, author's rebuttal letters, and editorial decisions. The peer review history covers all data from the initial submission to the final acceptance. With such abundant provision of data, the whole peer review process can be reproduced.

It should be noted that papers on PeerJ are published together with their original manuscripts available in PDF format, presenting the initial state of the papers as they are first submitted. By contrast, most of the other journals that do provide public access to the peer review data only present the published version of their papers, whereas the original manuscripts are usually made confidential. Without the original manuscripts, many open peer review-related studies will be rendered impossible. For example, it would be meaningless to study the review comments based on the revision. In the revision, the advice from the review comments has been taken and the related problems in the original manuscripts have been addressed. Without a reliable comparison with the original manuscripts, the effectiveness of the review comments will not be fully understood and the review comments will be rendered less effective for studies on open peer review.

On PeerJ, the review comments and the meta-reviews are both provided in plain text. This way no information will be lost during the conversion, keeping the original text of the reviews entirely intact. In addition, author's rebuttal letters to the review comments are provided in PDF, DOC, or DOCX format. The completeness of these data unlocks great possibilities for future studies on open peer review. Moreover, the review comments on PeerJ are well-structured with four segments: Basic Reporting, Experimental Design, Validity of the Findings, and Additional Comments. These four segments are very helpful as they can be used directly for labeling.

\begin{figure}[htb]
	\centering
	\includegraphics[width=0.8\textwidth]{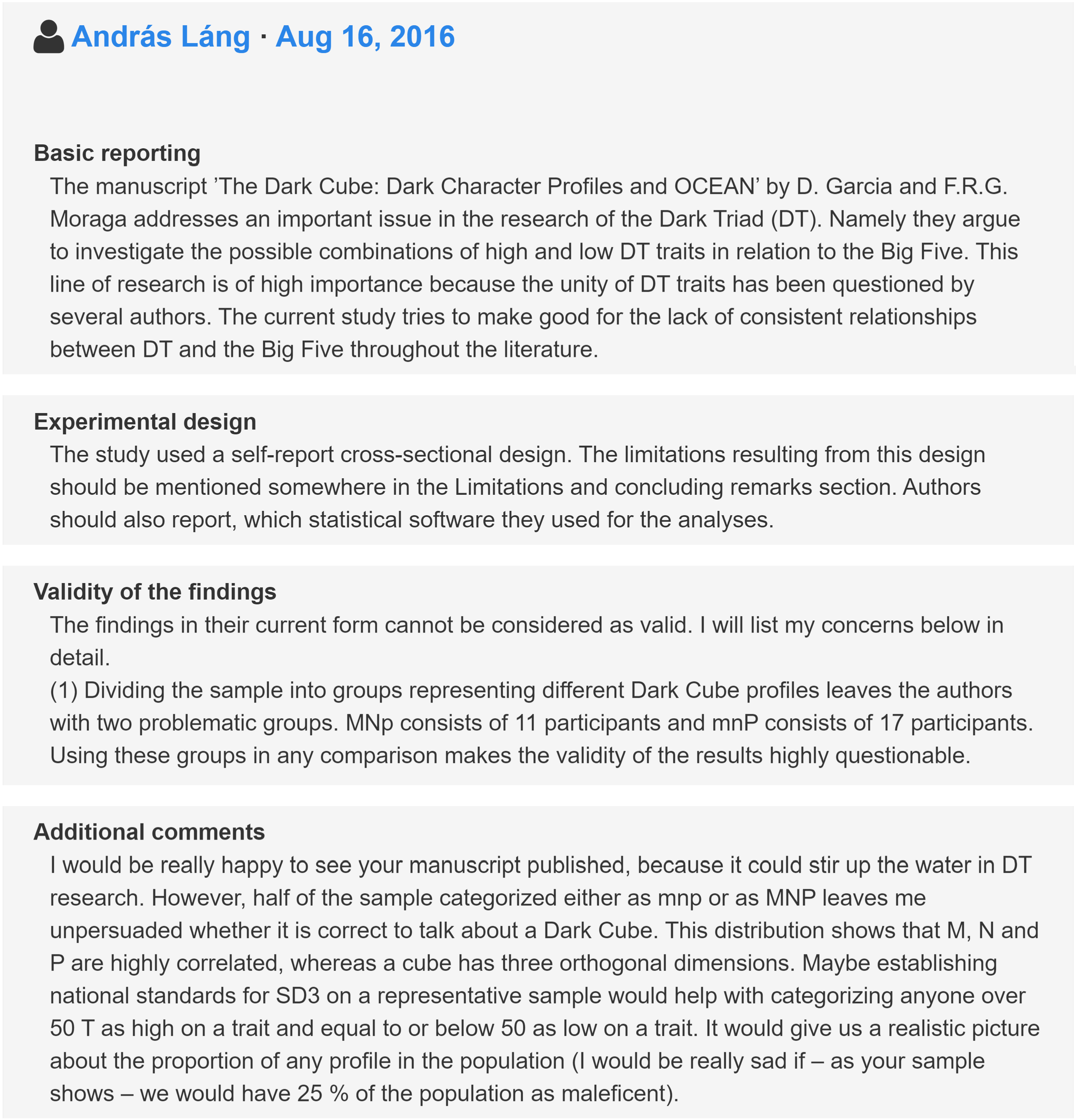}
	\caption{A review comment example on PeerJ (screenshotted from https://peerj.com/articles/3845/reviews/ with cropping to save space)}
	\label{fig:peerj-review-page}
\end{figure}

Apart from data transparency and completeness, what makes PeerJ stand out is the openness that comes with the Creative Commons Attribution License. With this highly accommodating license, peer review data on PeerJ are allowed to be collected, remixed, adapted, and reused as the foundation for new studies. Fig.~\ref{fig:peerj-review-page} shows how the review comment is provided on the web page of PeerJ.

\subsection{Construction}

A special crawler is designed to collect available complete peer review data for every paper on PeerJ. First, the metadata such as paper ID, paper title, paper discipline, and paper subject are collected. Next, the review comments, meta-reviews, and editorial decisions are crawled. Lastly, files of the initial submissions, revisions, and author's rebuttal letters are downloaded.

MOPRD is available in two versions. One is the Native MOPRD that contains all the files crawled from the web pages in HTML format as well as the PDF, DOC, and DOCX files downloaded. It is provided as an accommodating version for users to process and convert the open peer review data as needed. The other version is the Processed MOPRD. In this version, all the useful information of the HTML files in the Native MOPRD is parsed and provided in JSON format. The PDF manuscripts are converted into XML files using GROBID~\citep{lopez-grobid-2009}. Rebuttal letters in PDF format are parsed into TXT files by pdftotext.\footnote{https://poppler.freedesktop.org/.} Rebuttal letters in DOC and DOCX format are also parsed into TXT files by LibreOffice Writer.\footnote{https://www.libreoffice.org/discover/writer/.} Both the Native MOPRD and the Processed MOPRD can be downloaded from our website.\footnote{http://www.linjialiang.net/publications/moprd.} MOPRD is a long-term project that will be updated and extended to cover more journals in more disciplines when adequate resources are available.

\subsection{Properties}

In total, 6,578 papers are collected in MOPRD.\footnote{The data were collected on Aug 16, 2022.} The statistical features of MOPRD are listed in Table~\ref{tab:moprd-feature}.

\begin{table}[htb]
    \centering
	\caption{Statistical features of MOPRD}
	\label{tab:moprd-feature}
		\begin{tabular}{|l|l|}
			\hline
			Statistical feature                    &    Value  \\
			\hline
			Total papers                           &    6,578  \\
			\hline
			Total review comments                  &    22,483  \\
			\hline
			Total rebuttal letters                 &    11,213  \\
			\hline
			Mean review rounds per manuscript      &    2.7  \\
			\hline
			Mean reviewer number per manuscript    &    2.4  \\
			\hline
			Mean v1 manuscript word count          &    5,434  \\
			\hline
			Mean v2 manuscript word count          &    6,015  \\
			\hline
			Mean v1 review comment word count per reviewer   &    636  \\
			\hline
			Mean v2 review comment word count per reviewer   &    224  \\
			\hline
			Mean meta-review word count            &    129  \\
			\hline
		\end{tabular}
\end{table}

\begin{figure}[htb]
	\centering
	\includegraphics[width=0.9\textwidth]{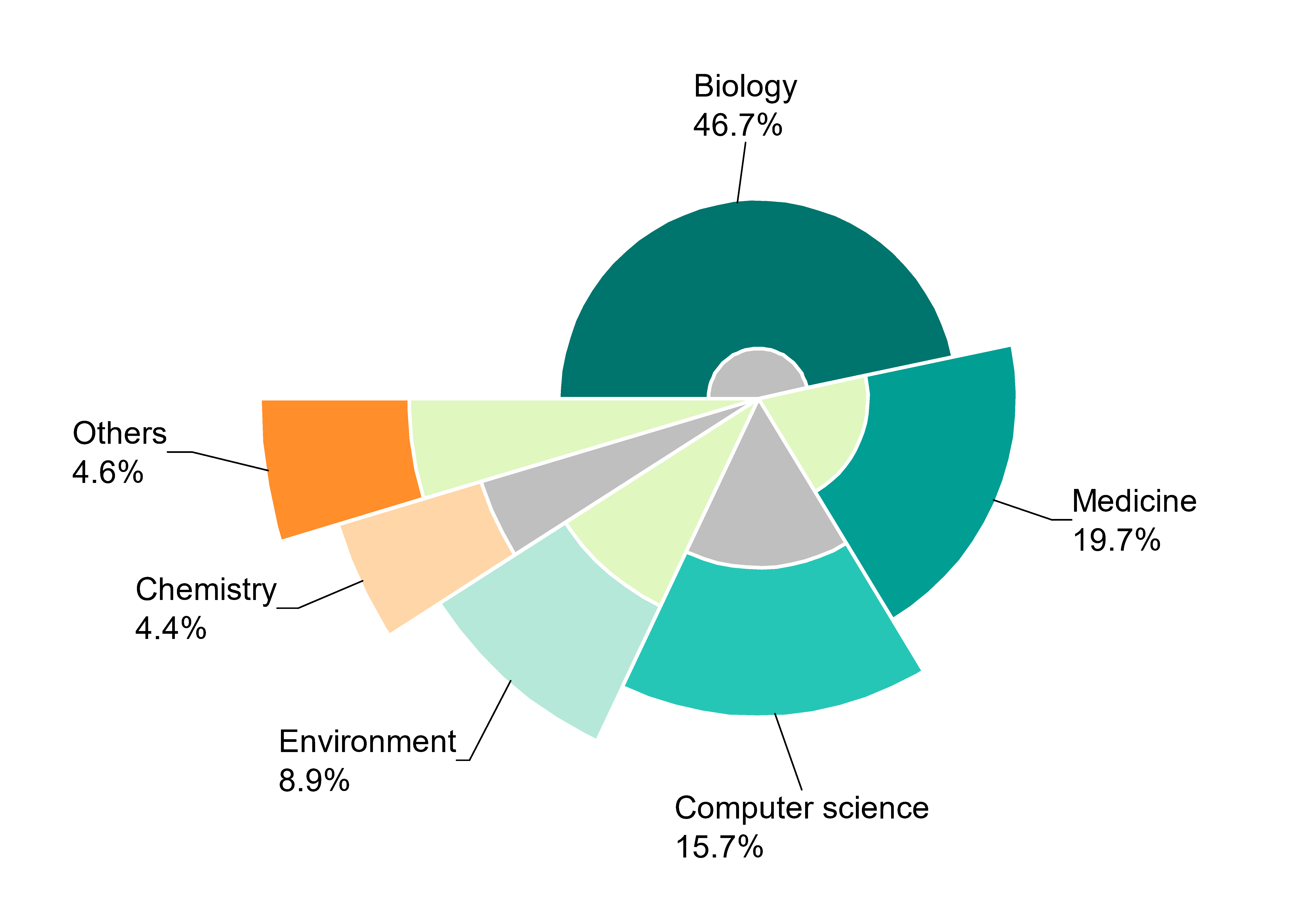}
	\caption{Disciplines of papers of MOPRD (one paper may fall into more than one discipline)}
	\label{fig:discipline-distribution}
\end{figure}

Disciplines to which these papers belong are presented in Fig.~\ref{fig:discipline-distribution}.

\section{Review comment generation}

As pointed out in \citet{lin-automated-2023}, ASPR can bring about remarkable benefits, including improving efficiency and saving resources for both the academic and publishing communities. Review comment generation is one of the key steps in achieving ASPR. Datasets play a highly important role in the realization of review comment generation. MOPRD as a large-scale dataset of review comment is highly suitable for the task of review comment generation.

\subsection{Method}

In this subsection, we elucidate the modular guided method for review comment generation. In essence, review comment generation is a task of text generation. The standard expression adopted in this research is as follows:
\begin{align*}
output = Postproc(Generate(Preproc(input)))
\end{align*}

where $input$ is a PDF manuscript that needs to be reviewed; $Preproc$ is the preprocessing method used for processing the PDF manuscript; $Generate$ is the generation method used after the preprocessing; $Postproc$ in our study represents a postprocessing method to integrate the results of $Generate$ into a whole review comment and convert it into a suitable format as users require; $output$ is the final generated review comments.

There are two main problems besetting the studies of review comment generation, with input length being one and labeled data being the other.

Firstly, in terms of input length, papers are usually thousands of words, and the number can extend to above 10,000 for some papers. However, the maximum input length of the previous generation of deep learning generative models, such as BART\citep{lewis-bart-2020}, is only 1,024 tokens. A token is a smaller unit for the separated pieces of text. Generally, 1 token is equal to approximately 0.75 words. So the text length of papers is beyond the processing power of traditional generative models. In response to this problem, \citet{yuan-can-2022} use an extract-then-generate method that reduces the text length to within the processing limits of 1,024 tokens. This method does address the problem of length limits for text generation, but it comes with a cost because an enormous amount of information in the original text is lost in the extraction processing. This way, the generated review comments will not be able to present thorough feedback on the paper. In response to this lingering problem, special attention mechanisms are designed to ease the input limits for the processing of long-text input. Some examples include Big Bird~\citep{zaheer-big-2020}, Longformer-Encoder-Decoder (LED)~\citep{beltagy-longformer-2020}, and LongT5~\citep{guo-longt5-2022}.

However, the input limits still exist as the maximum length of the input text only reach 16,384 tokens. In addition, the performance of attention mechanisms will be compromised more and more as the input length increases. Because the longer the input sequence, the more difficult for attention mechanisms to capture the full text features. To counteract this drawback, we design the strategy of modular generation by segmenting the input manuscripts based on the structure of the review comments.

Secondly, in review comment generation, labeled data plays a vital role in understanding the review comments. Generating a review comment that provides feedback for the paper from different perspectives requires a huge amount of work in extracting, locating, and labeling the data. Currently, labeling is best conducted by humans, but manual labeling is too costly for most projects in the field. What adds to the complexity of labeling in review comment generation is the multidisciplinary knowledge involved. Manual labeling the multidisciplinary data can only be made possible with experts from different fields. With PeerJ being the data source, the review comments in MOPRD are already being labeled as they are originally composed of four different segments. This makes MOPRD a dataset of great convenience for this task. Further, auto-labeling methods are proposed in this paper for automatic Questions and Proposals labeling. Auto-labeling can spare studies from costly handwork and is capable of multidisciplinary data.

In the following part of this section, we describe the preprocessing, preparation for generation, generation, and postprocessing steps in detail based on the expression above for review comment generation.

\subsubsection{Preprocessing}

The first step of preprocessing is to convert the input of PDF file into structured text. We use GROBID to convert PDF files into XML files. In the converted XML files, the manuscripts are presented following the original section structure. This supports our further processing of the manuscripts. The second step of preprocessing is to divide the text into three subsets of $t_{sum}$, $t_{mr}$, and $t_{full}$. Each subset is closely related to a certain segment of the review comments.
\begin{align*}
t_{sum}, t_{mr}, t_{full} = Preproc(input)
\end{align*}

$t_{sum}$ is composed of abstract, introduction, background, related work, future work, conclusion of a manuscript. It is a general summary of the manuscript. $t_{mr}$ consists of the sections of methods and results, which is the core contents of a manuscript. $t_{full}$ is the full text of a manuscript. It contains all the information and details of the manuscript.

A SciBERT-based~\citep{beltagy-scibert-2019} classification method, instead of keyword matching, is adopted to segment the manuscripts into three subsets of $t_{sum}$, $t_{mr}$, and $t_{full}$ by their section titles. Keyword matching is unreliable as different authors have their own wording and preferred layout of a paper. 120 manuscripts (20 per discipline) in MOPRD are randomly selected and parsed with GROBID to collect their section titles. We manually label these section titles into three classes of summary, methods \& results, and others. The labeled data are used to train and evaluate the classification model. The trained model achieves an accuracy of 0.986 and a weighted F1-score of 0.986.

\subsubsection{Preparation for generation}
\label{subsubsec:preparation-generation}

Several models are available for text generation with long-text input. Taking into consideration the performance and hardware requirement of different models, LED~\citep{beltagy-longformer-2020} is chosen as the pre-trained model for the task of review comment generation. LED, a variant of Longformer, is a seq2seq model supporting long-text input. LED adopts the mechanism of sliding window attention to reduce the spatio-temporal complexity to a linear relationship in sequence length for better performance in long-text modeling, which extends the input length limit to 16,384 tokens. It should be noted that what pre-trained model to use is not the key point of our modular guided method. There are plenty of alternative options when it comes to pre-trained models for text generation in our modular guided method, provided that they have the capacity of handling long-text input with supportive hardware.

The LED model is fine-tuned with the data collected in MOPRD before used in actual generation. Fine-tuning is a deep neural network training technique for transfer learning. The weights of a pre-trained model are taken as initial weights for a new model to be trained on a dataset smaller than the training dataset of the pre-trained model for weight modifications. The output layers can also be customized for specific tasks. In Fine-tuning, learned features are transferred to benefit downstream tasks, sparing them from large-scale training and greatly saving time, effort, and other resources. Fine-tuning has been proved to be highly advantageous to improve text generation performance.

Building high-quality fine-tuning datasets is of great importance in the fine-tuning of pre-trained models for text generation. A fine-tuning dataset for text generation consists of two parts: source text and target text. The purpose of fine-tuning is to adjust the pre-trained model to generate text resembling the target text as closely as possible based on the given source text. We build five different fine-tuning datasets to fine-tune five different models for generation. The expressions to achieve this are given as follows:
\begin{align*}
& Gen_{basic} = FineTune_{LED}(T_{sum}, R_{basic})  \\
& Gen_{ef} = FineTune_{LED}(T_{mr}, R_{ef})  \\
& Gen_{ques} = FineTune_{LED}(T_{full}, R_{ques})  \\
& Gen_{propos} = FineTune_{LED}(T_{full}, R_{propos})  \\
& Gen_{addl} = FineTune_{LED}(T_{full}, R_{addl})
\end{align*}

$FineTune_{LED}$ means using the first parameter as source text and the second parameter as target text to fine-tune the LED model. $T_{sum}$ is the set of all $t_{sum}$. $T_{mr}$ is the set of all $t_{mr}$. $T_{full}$ is the set of all $t_{full}$. $t_{sum}$, $t_{mr}$, and $t_{full}$ are all from the initial submissions of the papers in MOPRD. $R_{basic}$ is the set of Basic Reporting from all the review comments. $R_{ef}$ is the set of Experiments \& Findings with concatenated Experimental Design and Validity of Findings from all the review comments. $R_{addl}$ is the set of Additional Comments from all the review comments.

$R_{ques}$ is the set of Questions from all the review comments. It is a set of all $r_{{ques}_{i}}$. $r_{{ques}_{i}}$ is all the question sentences in $r_{{whole}_{i}}$. $r_{{whole}_{i}}$ is defined as $i$-th whole review comment with Basic Reporting, Experimental Design, Validity of Findings, and Additional Comments concatenated together. $r_{{propos}_{i}}$ is all the proposal sentences in $r_{{whole}_{i}}$. $R_{propos}$ is the set of Proposals with all $r_{{propos}_{i}}$.

$r_{{ques}_{i}}$ and $r_{{propos}_{i}}$ are not native data of PeerJ. They are obtained by the authors through an auto-labeling method. First, $r_{{whole}_{i}}$ is split into sentences using NLTK~\citep{loper-nltk-2002}. After all the sentences are identified, sentences with a question mark are collected for $r_{{ques}_{i}}$. Sentences containing keywords like ``suggest'', ``ought to'', ``should'', which indicate the sentence to be a proposal, are gathered for $r_{{propos}_{i}}$. Moreover, POS tagging is conducted on the sentences using NLTK. A sentence that starts with a verb and ends with a period is recognized as an imperative sentence. Such sentences are also collected into $r_{{propos}_{i}}$.

Furthermore, extra processing is performed for all the review comments. Our model is not trained to cope with figures and tables, therefore review comments on figures and tables in the dataset will affect the performance of the generation model. In some cases, undesired review comments on figures and tables can be generated. To address this problem, sentences with ``figure'', ``fig.'' or ``table'' are removed from the review comments. As the review comments are collected from PeerJ, some are written especially for publication on PeerJ, so sentences containing ``PeerJ'' are also filtered out for greater accommodation.

The review comment labels provided in MOPRD enable the fine-tuning with different subsets of the manuscripts as source text and their related segments of the review comments as target text. In addition to the original labeled segments of review comments, two more segments of Questions and Proposals are created through auto-labeling. In peer review, it is very common for reviewers to raise questions and proposals in order to settle confusion in the manuscripts. It helps authors improve their manuscripts with clarity and completeness. However, efforts are yet to be given to these two types of data in review comment generation. Therefore, in this study, we take the pioneering step to tap into the data of questions and proposals for review comment generation.

These five models are fine-tuned separately. The general summary of a manuscript is related to the segment of Basic Reporting in the review comment. Therefore, $T_{sum}$ is fed as the source text, and $R_{basic}$ is fed as the target text. Both are input into one model for fine-tuning to obtain $Gen_{basic}$. Similarly, the method and the result sections of a manuscript are related to the segment of Experimental Design and the segment of Validity of Findings in the review comment. Thus, $T_{mr}$ is used as the source text, and $R_{ef}$ serves as the target text. These are input into another model for fine-tuning to produce $Gen_{ef}$. The segments of Questions, Proposals, and Additional Comments in the review comment pertain to the full content of a manuscript. Hence, $T_{full}$ serves as the source text for the target texts of $R_{ques}$, $R_{propos}$, and $R_{addl}$, respectively. These are input into three separate models for fine-tuning to generate $Gen_{ques}$, $Gen_{propos}$, and $Gen_{addl}$.

\subsubsection{Generation}

After the fine-tuning, five different models, i.e.\ $Gen_{basic}$, $Gen_{ef}$, $Gen_{ques}$, $Gen_{propos}$, and $Gen_{addl}$ are built to generate five different modules of review comment with a guided approach. Our generation expression is given below:
\begin{align*}
Generate=\left\{
\begin{array}{ll}
	& Gen_{basic}(p_{basic}, t_{sum})  \\
	& Gen_{ef}(p_{ef}, t_{mr})  \\
	& Gen_{ques}(p_{ques}, t_{full})  \\
	& Gen_{propos}(p_{propos}, t_{full})  \\
	& Gen_{addl}(p_{addl}, t_{full})  \\
\end{array} \right.
\end{align*}

where $p_{basic}$, $p_{ef}$, $p_{ques}$, $p_{propos}$, and $p_{addl}$ are the prefixes fed respectively to the models to guide the generation of Basic Reporting, Experiments \& Findings, Questions, Proposals, and Additional Comments.

These prefixes are randomly selected from their corresponding predefined sets. A predefined set comprises many phrases or sentences with similar meanings. For example, the prefixes of $p_{basic}$ include ``In this paper, the authors proposed'' and ``In this study, the author explored''; the prefixes of $p_{ef}$ contain ``The experimental design is'' and ``The method of this paper''; the prefixes of $p_{ques}$ cover ``I have several questions:'' and ``Some questions are raised.''; the prefixes of $p_{propos}$ contain ``I do have some important recommendations for improving the paper.'' and ``I have some following suggestions:''; the prefixes of $p_{addl}$ include ``There are still some comments for the authors to substantively revise the manuscript.'' and ``Additionally, I have some other comments:''.

In general, contextual coherence and schematic structure are missing in the review comments generated by pre-trained models. But in our modular guided method, each model is guided to generate specific contents for the corresponding module. The generated review comment can thus follow the structure of real-life review comments with higher completeness.

\subsubsection{Postprocessing}

There are two main functions of $Postproc$. One is to concatenate the outputs of the five models in $Generate$ to be a whole review comment. The other is to convert the output into a reader-friendly format.

\subsection{Discussion}

\begin{figure}[htb]
	\centering
	\includegraphics[width=0.9\textwidth]{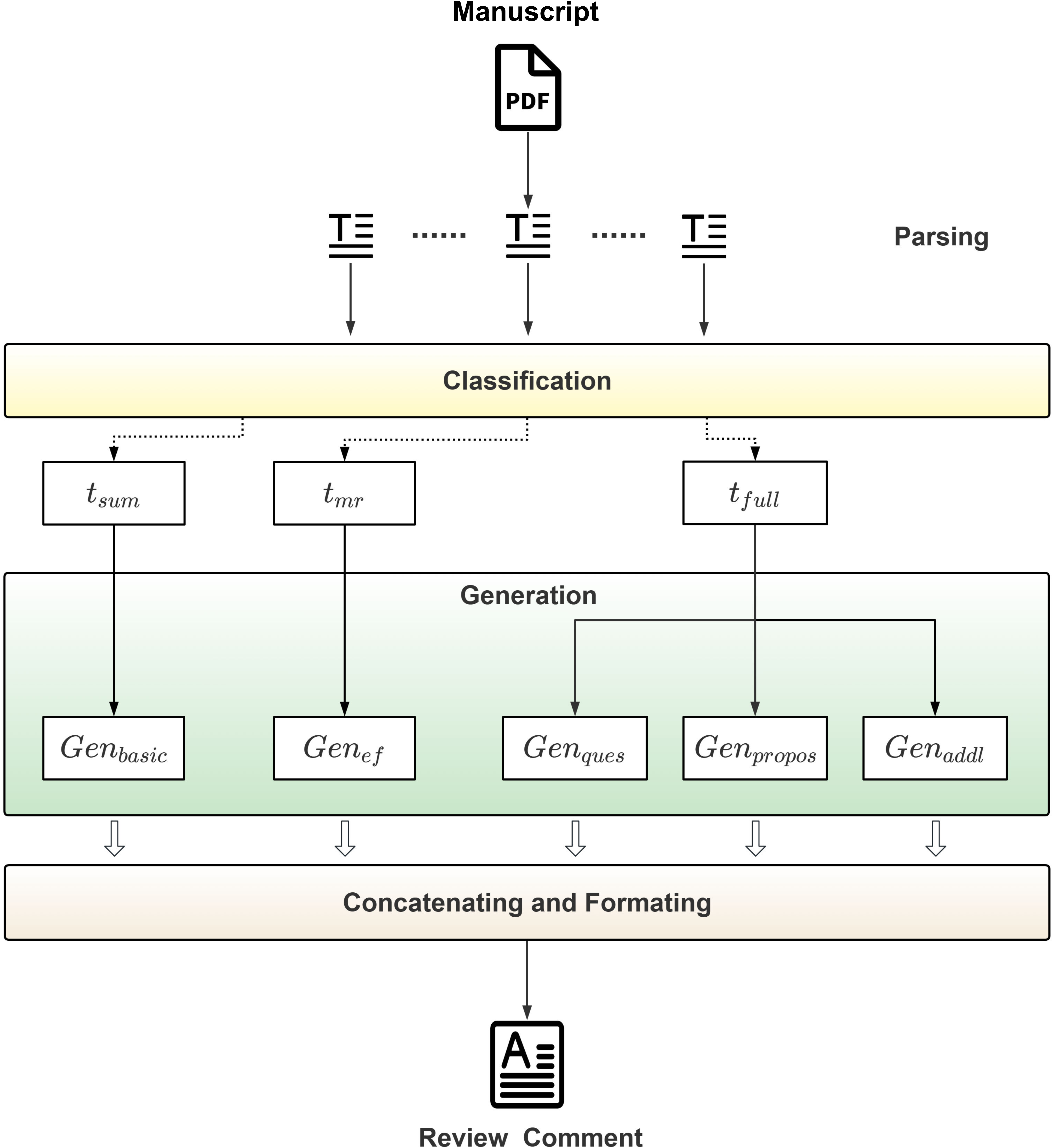}
	\caption{Illustration of our modular guided generation method}
	\label{fig:illustration-generation-method}
\end{figure}

The whole process of review comment generation with our modular guided method is illustrated in Fig.~\ref{fig:illustration-generation-method}.

Our method stands out with two special designs, the modular design and the guided design. First, the modular design facilitates the handling of long manuscripts and the generation of longer and more comprehensive review comments. With the use of LED and the approach of modular manuscript segmentation, long manuscripts can be processed effectively. Furthermore, the approach of modular generation is adopted. With five generation models working together, the final generated review comment can be maximized to be five times the length of a traditional model. Second, through the guided design, the generated review comment of each module is appropriately guided. By combining with the modular design, a basic structure is established for the overall generated content, enhancing its logical coherence and making it more reader-friendly.

Our modular guided method is developed based on the review comments of PeerJ but, nonetheless, it is an accommodating method that can be reused for most papers. The review comments of PeerJ are composed of four segments of Basic Reporting, Experimental Design, Validity of the Findings, and Additional Comments. Apart from PeerJ, most papers also have their review comments in a similar, if not the same, structure. Therefore, papers other than PeerJ can use our method equally. For papers with a special review comment structure, our method is also applicable when provided with a related labeled data source.

Concerns might be raised about the multidisciplinary bias of our method. We notice that papers in different disciplines have various writing styles and focuses. For example, in the field of AI, papers focus primarily on the originality or novelty of the proposed methods or models, while importance is attached more to experimental designs and results in the field of medicine. However, this situation does not invalidate our method. Nowadays, interdisciplinary research is on the rise~\citep{van-interdisciplinary-2015}, and research methods from different disciplines also continue to merge. With enough data and appropriate selection and balancing strategies, it is consequently possible to learn enough features across disciplines in this interdisciplinary trend. Additionally, \citet{soltau-neural-2017} have proved that a simple end-to-end model can produce better results when trained with large-scale data, compared with a complex model trained with small-scale data. Therefore, large-scale datasets can be a performance booster for AI research. MOPRD, as the largest multidisciplinary open peer review dataset by far, is hopeful to continually elicit more and better data from the research community. At this moment, we just evaluate our method on MOPRD and provide the baseline. Once enough data are collected, the multidisciplinary bias can be effectively solved.

\subsection{Experiments}

In this subsection, we present the technical details of our modular guided method, conduct a sanity test, compare our method with the SOTA review comment generation method of \citet{yuan-can-2022} with MOPRD, and assess the model performance between training with single-discipline data and multi-discipline data.

In the experiments, only the initial submissions and their corresponding review comments in the MOPRD are used. One reason is that review comments on the initial submissions are generally longer with more thorough contents. The other reason is that some review comments on revisions invoke the contents in the revisions and the author's rebuttal letters. These data will add to the complexity of review comment generation. Manuscripts less than 2,000 words and review comments less than 100 words are filtered out. A manuscript and one of its review comments are set as a record. The final dataset for the experiments consists of 10,998 records of manuscript-comment pairs. These records are divided randomly into the training set, validation set, and test set at the ratio of 8:1:1.

We use the Transformers~\citep{wolf-transformers-2020} framework and the NVIDIA A40 (48GB GPU memory) for the experiments. The conventional beam search is not used in our method, because beam search may lead to the generation of repetitive contents among different papers. Especially, the sentences in the generated review comments might share the same sentence starters. Instead, we use Top-P and Top-K sampling~\citep{fan-hierarchical-2018} jointly with p being 0.92 and k being 50. This combination promises more diversified and more human-like comments. N-gram penalty is applied to minimize the generation of repetitive words~\citep{klein-opennmt-2017,paulus-deep-2018} with no\_repeat\_ngram\_size being 3. Each model can be fine-tuned within 24 hours.

\subsubsection{Sanity test}

\begin{figure}[htb]
	\centering
	\includegraphics[width=0.9\textwidth]{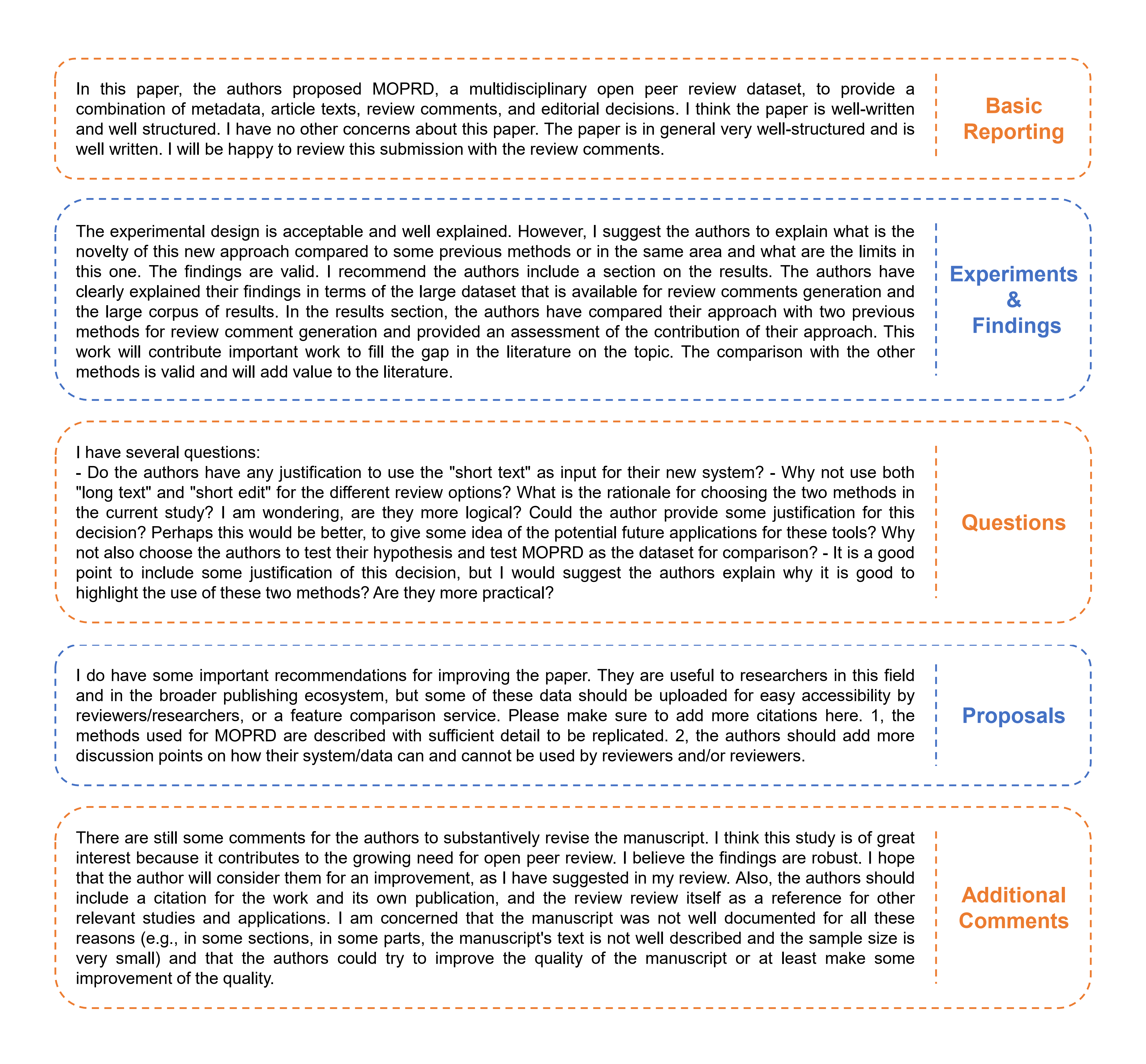}
	\caption{The review comment generated with our modular guided method of a drafted version of the current paper}
	\label{fig:sanity-test-moprd-review}
\end{figure}

We perform a sanity test on our method by generating a review comment for a draft version of the current paper ``MOPRD: A Multidisciplinary Open Peer Review Dataset''. The draft version is a near-finished version before the final version and the section of Sanity Test is deleted in the sanity test on our method. The generated review comment is presented in Fig.~\ref{fig:sanity-test-moprd-review} without any alteration. Some revisions have been made to the draft version based on the review comment.

As shown in Fig.~\ref{fig:sanity-test-moprd-review}, the generated review comment does address the focal points of each target module, while also being pertinent to the content of the paper itself. Moreover, there is no conventional rigidity, which is often associated with template-based generation, shown in the generated review comment. Given the combined utilization of Top-P and Top-K sampling for content selection during generation, a small portion of the output may exhibit grammatical errors and inexplicable phrasing. This is the focus for future improvement.

\subsubsection{Comparison of generation methods over automatic metrics}

Then, we use ROUGE~\citep{lin-automatic-2003,lin-rouge-2004} and BARTScore~\citep{yuan-bartscore-2021} as the automatic metrics to evaluate the performance of different generation methods.

ROUGE, denoted Recall-Oriented Understudy for Gisting Evaluation, is a set of evaluation metrics used for summarization and other similar text generation tasks. It works by measuring the number of overlapping units between the generated text and the reference text. With the evaluation of ROUGE, greater performance is indicated by more overlapping units. Several commonly used ROUGE metrics are ROUGE-1, ROUGE-2, ROUGE-L, and ROUGE-Lsum. Higher ROUGE scores mean better performance.

BARTScore is a relatively new metric that uses text generation probability to evaluate the quality of the generated text. BARTScore calculates the probability of transforming the generated text into the reference text. Pre-trained by seq2seq models, BARTScore is able to evaluate the generated text from different aspects, including fluency, accuracy, and validity. BARTScore is a negative value with a lower absolute value suggesting better performance.

The extract-then-generate method of review generation proposed by \citet{yuan-can-2022} is used for comparison. The same training set, validation set, and test set built on MOPRD are used both for our method and the extract-then-generate method. The compared method is conducted by carefully following the descriptions in the paper that introduces it.

Furthermore, to fully test the effectiveness of our method, we conduct two extra experiments to validate the two special designs in our method. One is the segmentation-less experiment for validation the modular design. In the fine-tuning of the segmentation-less method, the full text of the manuscripts is fed as source text, and the whole review comments are fed as target text. The expressions are as follows:
\begin{align*}
Gen_{segless} = FineTune_{LED}(T_{full}, R_{whole})
\end{align*}

\begin{align*}
Generate_{segless}=Gen_{segless}(p_{empty}, t_{full})
\end{align*}

where $R_{whole}$ is a set of all $r_{{whole}_{i}}$ (described in Sect.~\ref{subsubsec:preparation-generation}); $p_{empty}$ is an empty string to indicate that no prefix is given for guiding.

Another is the modular non-guided experiment for validation the guided design. In the modular non-guided method, the prefixes of each generation model are removed, meaning that prefixes are not used to guide the generated content. Apart from this, the setup of this experiment remains the same as that of the modular guided experiment.

The results of these experiments are listed in Table~\ref{tab:review-automatic-metric-comparison}.

\begin{table}[htb]
    \centering
	\caption{Comparison of review comment generation methods over automatic metrics}
	\label{tab:review-automatic-metric-comparison}
	\resizebox{0.96\textwidth}{!}{
		\begin{tabular}{|l|l|l|l|l|l|}
			\hline
			Method										   &   ROUGE-1          &   ROUGE-2         &   ROUGE-L          &   ROUGE-Lsum       &   BARTScore   \\
			\hline
			Extract-then-generate~\citep{yuan-can-2022}    &     35.59          &     8.57          &   \textbf{16.10}   &     33.99          &     $-$4.19   \\
			\hline
			Segmentation-less                              &     37.71          &     9.00          &     15.37          &     36.22          &     $-$4.16   \\
			\hline
			Modular non-guided                             &     38.70          &     8.88          &     15.43          &     37.24          &     $-$4.16   \\
			\hline
			Modular guided (proposed)                      &   \textbf{39.01}   &   \textbf{9.10}   &     15.49          &   \textbf{37.53}   &   \textbf{$-$4.15}   \\
			\hline
		\end{tabular}
	}
	\begin{tablenotes}
		\item \footnotesize{Bold values indicate the best value in a column}
	\end{tablenotes}
\end{table}

As displayed in Table~\ref{tab:review-automatic-metric-comparison}, our proposed method outperforms all other methods over all metrics except ROUGE-L in the comparison with the extract-then-generate method. The extract-then-generate method has the highest ROUGE-L score among all methods. This is not surprising as the extract-then-generate method uses beam search with the number of beams set to 4. This design tends to generate phrases with the four words of the highest probability, hence the high ROUGE-L score. But beam search entails repetition problems in its generated text. The segmentation-less method delivers greater performance than the extract-then-generate method over all other metrics. This reveals that using full text rather than extracted summary as input is more effective in feature representation. The modular non-guided method surpasses the segmentation-less method in most metrics. This indicates that the modular design is capable of better performance. However, without the guided design, the modular design can not make the most of it. The modular design and the guided design need to work together to achieve the best results. That is why the modular non-guided method fails to produce better results than the segmentation-less method in all metrics. Overall, the modular guided method surpasses both the segmentation-less method and the modular non-guided method over all metrics. This validates the effectiveness of our proposed method.

\subsubsection{Comparison of generation methods over human evaluation}

After testing our method with the two automatic evaluation metrics of ROUGE and BARTScore, its real-world quality is put to test with human evaluation. 10 researchers from the disciplines of biology, chemistry, computer science, environment, and medicine are invited for human evaluation of the generated review comments. Each researcher is assigned to evaluate the same generated review comments of two manuscripts selected based on their research fields. In all, 10 manuscripts with a total of 20 review comments generated by two different methods are evaluated as samples. Each review comment is independently scored by two different researchers. The researchers score the generated review comments based on structure \& logicality, specialty \& accuracy, and usefulness \& inspiration. A five-point rating scale is used, with higher points indicating better results. The results are presented in Table~\ref{tab:review-human-evaluation-comparison}.

\begin{table}[htb]
    \centering
	\caption{Comparison of review comment generation methods over human evaluation}
	\label{tab:review-human-evaluation-comparison}
	\resizebox{0.96\textwidth}{!}{
		\begin{tabular}{|l|l|l|l|l|}
			\hline
			Method                                         &    Structure \& logicality    &    Specialty \& accuracy    &    Usefulness \& inspiration  \\
			\hline
			Extract-then-generate~\citep{yuan-can-2022}    &   2.85                        &   2.75                      &   2.65   \\
			\hline
			Modular guided (proposed)                      &   \textbf{3.45}               &   \textbf{3.30}             &   \textbf{3.40}   \\
			\hline
		\end{tabular}
	}
	\begin{tablenotes}
		\item \footnotesize{Bold values indicate the best value in a column}
	\end{tablenotes}
\end{table}

As shown in Table~\ref{tab:review-human-evaluation-comparison}, our method surpasses the compared method over all metrics. In addition, the scores of our method are all above 3 points. Compared to all other scores, our method has the highest score of 3.45 for structure \& logicality. This occurs naturally with our special modular guided design. In terms of specialty \& accuracy, the rating is relatively lower. Our following studies are aiming at improving performance in this aspect. Both our method and the compared method are capable of generating useful and inspiring review comments, which is the core of review comment generation. This shows that ASPR is to be achieved as a realistic goal. Review comment generation can benefit all parties in the academic and publication communities, including editors, reviewers, and authors.

\subsubsection{Single-discipline training data VS multi-discipline training data}

Here, an experiment is conducted to explore how training data affect model performance in review comment generation. Two models are trained under the same settings but with different training data, one with single-discipline data and another with multi-discipline data. The segmentation-less generation method is used to control the influence of other factors.

As displayed in Fig.~\ref{fig:discipline-distribution}, MOPRD mainly consists of data from five disciplines: biology, chemistry, computer science, environment, and medicine. Data from the discipline of biology are used for single-discipline training as biology is the largest discipline with the most data in MOPRD. Data from the disciplines of biology, chemistry, environment, medicine, and other disciplines with smaller data sizes are used for multi-discipline training. The biology data used for single-discipline training are not overlapped with those biology data used for multi-discipline training. The ratio of single-discipline training data to multi-discipline training data is 1:1. To resolve overlapping between the training data and test data, the test data is composed of the discipline of computer science only as no computer science data are used as the training data. The results are listed in Table~\ref{tab:single-multi-discipline-comparison}.

\begin{table}[htb]
    \centering
	\caption{Comparison of single-discipline and multi-discipline models in review comment generation}
	\label{tab:single-multi-discipline-comparison}
	\resizebox{0.96\textwidth}{!}{
	\begin{tabular}{|l|l|l|l|l|l|}
		\hline
		Data                                     &   ROUGE-1          &   ROUGE-2         &   ROUGE-L          &   ROUGE-Lsum        &   BARTScore   \\
		\hline
		Single-discipline                        &   29.97            &   7.52            &   14.07            &   28.88             &   $-$4.25   \\
		\hline
		Multi-discipline                         &   \textbf{30.96}   &   \textbf{7.98}   &   \textbf{14.25}   &   \textbf{29.88}    &   \textbf{$-$4.23}   \\
		\hline
	\end{tabular}
	}
	\begin{tablenotes}
		\item \footnotesize{Bold values indicate the best value in a column}
	\end{tablenotes}
\end{table}

As displayed in Table~\ref{tab:single-multi-discipline-comparison}, it is indicated by all metrics that the multi-discipline model can deliver better performance in review comment generation than the single-discipline model. This proves that models trained with multi-discipline data are more capable of transferring the learned features in the generation of review comments for unknown disciplines. However, most existing review comment datasets are comprised either entirely or mostly of data from the discipline of computer science only. These datasets are inadequate for review comment generation for papers from other disciplines. But MOPRD as a multi-discipline dataset can rectify this inadequacy to generate proper review comments for papers from multiple disciplines.

It is also noted that there is performance inconsistency between the models in this experiment and the models in the previous experiments, which can be seen in Table~\ref{tab:review-automatic-metric-comparison}. Review comments generated in this experiment are found to show lower scores. One reason is that the models in this experiment are applied to generate review comments for an unknown discipline in the training. Another reason is that the size of training data for the models in this experiment is significantly smaller compared to that of the models in the previous experiments. The training data are halved in size in this experiment.

\section{Other applications}

MOPRD as a dataset of abundant peer review data has great potential for many other applications. In this section, we discuss some of the possible applications of MOPRD.

\subsection{Meta-review generation}

Meta-reviews are well-founded arguments written by editors to give explicit justification for the decision of a manuscript by reconciling all the review comments made by the reviewers. Meta-review generation can assist editors by providing an overview of all the review comments of a certain manuscript. In essence, meta-review generation is a multi-document summary task with several review comments as input and the meta-review as output.

Some studies~\citep{pradhan-deep-2021,shen-mred-2022} have been conducted before on this task with other datasets. Compared to those datasets, MOPRD stands out with its completeness of peer review data because the initial submissions and all the revisions of the papers are collected. These full peer review data enable the generation of meta-reviews in different rounds, and even using the content of manuscripts. The multi-document summary model of PRIMERA~\citep{xiao-primera-2022} can be used with MOPRD for this task. PRIMERA is a pre-trained multi-document representation model created for the task of summarization. In contrast to previous approaches that generate summaries with single-document representations and then aggregate them, PRIMERA uses multiple documents as input directly to obtain multiple representations for a one-stop generation of summaries. With PRIMERA, multiple documents are concatenated into one long text to be modeled by Longformer. An entity pyramid approach is also proposed along PRIMERA for better cross-document information aggregation with masked salient information.

\subsection{Editorial decision prediction}

Editorial decision prediction is a task of predicting the decision of acceptance, revision, and rejection of a manuscript based on the review comments~\citep{ghosal-deepsentipeer-2019,deng-hierarchical-2020}.\footnote{Apart from the review comments, the content of the manuscript itself is also used by some researchers for this task.} Editorial decision prediction can help editors by sorting out information into insights to facilitate decision making. In computer science, editorial decision prediction is considered a task of multi-document classification. It takes multiple review comments as input and eventually outputs acceptance, revision, or rejection predictions, with an intermediate process that may also include extracting core arguments from long text to corroborate decisions. The process of implementing this task relies on the multi-document representation, aggregation, and modeling process. From a more detailed perspective, it also depends on the analysis of the content of multi-documents, integration of key arguments, determination of sentiment tendencies~\citep{han-augmented-2019,huan-emotionally-2022}, and inference across documents. Research has been conducted on the sentiments shown in the review comments for editorial decision prediction~\citep{ghosal-sentiment-2019,ribeiro-acceptance-2021}. The review comments are analyzed to detect positive or negative attitudes of the reviewers towards a certain manuscript.

The advantage of using MOPRD for this task lies in its multidisciplinary data. Comparison can be conducted to study sentimental differences across the review comments from different disciplines, whereas previous analyses are not able to do a multidisciplinary comparison.

\subsection{Author rebuttal generation}

Author rebuttal generation can be a featured application of MOPRD, making use of the review comments, meta-reviews, and author's rebuttal letters collected in this dataset. For inexperienced authors, author rebuttal generation can provide guidance in the writing of their own rebuttal letters. A rebuttal letter is written by the authors of a certain manuscript to address the concerns of reviewers and editors. A typical author's rebuttal letter has a basic structure of three segments, i.e.\ the review comments (meta-review), the author's response, and the revision to the original manuscript. In a rebuttal letter, authors are taking up the opportunity to directly reply to the reviewers and editors in a point-by-point manner, defend aspects of work, eliminate contextual misunderstandings, and elaborate on the improvements to the manuscript.

Author rebuttal generation can be considered a question-answering application and is a task of multi-input text generation. The review comments, meta-review, and the reviewed manuscript are used as input and the author's rebuttal letter is used as the output. An author's rebuttal letter needs to make a point-by-point response to each review point sequentially. One review point may be raised specifically for a certain sentence, a certain paragraph, or a whole section of the manuscript. Therefore, in author rebuttal generation, it is crucial for this textual correspondence to be identified and reproduced. To achieve this, segmentation and classification play a vital role in the modeling of the text input. They help to establish the textual correspondence between the review points and the specific contents in the manuscript to guide the generation of the rebuttal response.

\subsection{Scientometric analysis}

Scientometrics is a complex of ``the quantitative methods of the research of the development of science as an informational process''~\citep{nalimov-measurement-1971}. Research on peer review is gathering momentum with more and more studies being conducted on peer review. To follow this trend, scientometrics, a quantitative science, is needed in this field to measure and analyze the process of peer review research. As the prerequisite of scientometric analysis, a large amount of data is indispensable. With the further development of open peer review and its gradual applications in various fields, research data and information about peer review process have become richer and clearer, and research topics have become more diverse and valuable. The formerly mysterious process of peer review can be further analyzed and interpreted using quantitative and data-based methods.

\citet{matsui-impact-2021} is an example of peer review scientometric analysis. Approximately 5,000 papers from three disciplines are collected and analyzed from two PeerJ journals. MOPRD has 20\% more data from a wider variety of disciplines than the data used in this study. Therefore, with MOPRD, more profound results can be achieved on peer review scientometric analysis.

\section{Conclusion}
\label{sec:conclusion}

In this paper, we introduce the Multidisciplinary Open Peer Review Dataset (MOPRD), which is, to our best knowledge, the first large-scale open peer review dataset of multidisciplinary data. This dataset contains paper metadata, multiple version manuscripts, review comments, meta-reviews, author's rebuttal letters, and editorial decisions of 6,578 papers. The data cover the entire peer review process from paper submission to publication. It brings great data diversity into peer review datasets as previously they are mainly composed of data from computer science. Based on MOPRD, we present a modular guided review comment generation method. This method is superior to the compared methods as measured by the automatic metrics of ROUGE and BARTScore. The method also completely surpasses the compared method in all aspects of structure \& logicality, specialty \& accuracy, and usefulness \& inspiration in human evaluation, with every aspect score above the mean. The proposed method delivers the best performance in the aspect of structure \& logicality among the three aspects. Our experiments also demonstrate that in review comment generation for unknown disciplines, models trained with multi-discipline data work better than models trained with single-discipline data. This is the reason why MOPRD is more suitable for real-life applications than the previous datasets. Last but not least, we also explore other applications of MOPRD including meta-review generation, editorial decision prediction, author rebuttal generation, and scientometric analysis. MOPRD is offered to the public to support future studies with its great potential.

\section*{Acknowledgements}
This work is partly funded by the 13th Five-Year Plan project Artificial Intelligence and Language of State Language Commission of China (Grant No. WT135-38). We appreciate Fangzhi Chen, Guantian Ding, Hongkun Fang, Jiabin Xue, Jingjing Wang, Jintao Guo, Li Lei, Ning Zhang, Zhou Xu, and Zhu Lin for their work in evaluating the review comments. Special and heartfelt gratitude goes to the first author's wife Fenmei Zhou, for her understanding and love. Her unwavering support and continuous encouragement enable this research to be possible.

\section*{Data availability}
The method of getting our dataset is provided within the paper.

\section*{Declarations}
\textbf{Conflict of interest} The authors declare that there is no conflict of interest regarding the publication of this paper.


\bibliography{sn-article}


\end{document}